\documentclass[conference,a4paper]{IEEEtran}
\IEEEoverridecommandlockouts
\usepackage{cite}
\usepackage{amsmath,amssymb,amsfonts}
\usepackage{graphicx}
\usepackage{textcomp}
\usepackage{xcolor}
\usepackage{caption}
\usepackage{siunitx}
\usepackage{algorithm}
\usepackage{algpseudocode}
\usepackage{balance}
\usepackage{subcaption}
\def\BibTeX{{\rm B\kern-.05em{\sc i\kern-.025em b}\kern-.08em
    T\kern-.1667em\lower.7ex\hbox{E}\kern-.125emX}}
\newcommand{\lb}{1097 }

\begin{document}

\title{Environmental Variation or Instrumental Drift? \\
A Probabilistic Approach to Gas Sensor Drift Modeling and Evaluation
\thanks{This research project has received funding from Eurostars project E115482 (CitySense). We would also like to acknowledge Senseair for the helpful technical discussions and Empa and Decentlab for providing the dataset.}
}
\author{
\IEEEauthorblockN{Cheng Yang,
                Gustav Bohlin,
                Tobias Oechtering}
\IEEEauthorblockA{
              KTH Royal Institute of Technology,  
              Stockholm,
              Sweden, 
              \{yangch,gbohlin,oech\}@kth.se}
}
\maketitle

\begin{abstract}
Drift is a significant issue that undermines the reliability of gas sensors. This paper introduces a probabilistic model to distinguish between environmental variation and instrumental drift, using low-cost non-dispersive infrared (NDIR) CO$_2$ sensors as a case study. Data from a long-term field experiment is analyzed to evaluate both sensor performance and environmental changes over time. Our approach employs importance sampling to isolate instrumental drift from environmental variation, providing a more accurate assessment of sensor performance. The results show that failing to account for environmental variation can significantly affect the evaluation of sensor drift, leading to improper calibration processes.
\end{abstract}

\begin{IEEEkeywords}
Sensor drift, probabilistic modeling, environmental variation, instrumental drift, NDIR CO$_2$ sensors, importance sampling.
\end{IEEEkeywords}

\section{Introduction}
\label{intro}
Gas sensors are vital in various applications, from environmental monitoring, industrial processes to healthcare and safety systems \cite{liu2012survey}. Their ability to provide real-time, accurate gas concentration measurements is essential for maintaining air quality, ensuring safety, and optimizing processes. However, one significant challenge for gas sensors is the phenomenon of drift.

Drift generally refers to the gradual changes in a quantitative characteristic that is assumed to be constant over time\cite{ziyatdinov2010drift}. It is a pervasive issue, particularly for low-cost gas sensors, which gained lots of attention recently because of their affordability and broad applicability. The causes of sensor drift, which manifests as undesired temporal changes in sensor's accuracy or error level, are often complex and context-specific, typically related to physical changes in the sensor (e.g., aging), environmental fluctuations, and cross-sensitivities.

Despite numerous studies addressing sensor drift \cite{artursson2000drift, padilla2010drift, wenzel2010online, vergara2012chemical, you2022time} (see review papers \cite{marco2012signal, maag2018survey}), there is a notable gap in modeling various factors contributing to drift. In this paper, we argue that drift sources  can be classified into environmental
variations and instrumental drift, which are specified
in mathematical models, and can be distinctly identified from
the data.

Several works have examined how environmental factors affect gas sensor performance. For instance, \cite{abidin2014temperature} explores gas sensor drift due to ambient temperature changes, \cite{robbiani2023physical} summarizes physical confounding factors affecting sensor responses, and \cite{maag2018survey} categorizes typical error sources in low-cost sensors and their calibration approaches. Additionally, there is a growing body of research utilizing probabilistic approaches to model sensor behavior, including sensor fault detection \cite{mehranbod2003probabilistic}, missing data modeling \cite{chai2021deep}, gas quantification \cite{monroy2013probabilistic}, and sensor fusion \cite{you2020belief}, etc. These probabilistic models provide robust frameworks for interpreting and predicting sensor performance.

Our contribution is twofold. First, we introduce a simple probabilistic model to distinguish between environmental variation and instrumental drift. We use a case study of a low-cost non-dispersive infrared (NDIR) CO$_2$ sensor to demonstrate this differentiation. Second, we provide a sampling-method-based solution for excluding the impact of environmental variation and evaluating instrumental drift in practical applications. 

By addressing these challenges, we aim to enhance the understanding of sensor drift and provide practical evaluation solutions, contributing to more reliable gas sensing and other sensor technologies facing similar issues.

\section{Probabilistic Drift Modeling}
\subsection{Probabilistic Modeling of NDIR Sensor}
NDIR sensors use a broadband light source without a diffraction grating or prism. Light passes through the gas in the optics cell, then through a narrow-band filter, and reaches the infrared detector. By measuring the detected infrared light intensity, the gas concentration can be calculated.

We consider three physical quantities in the measurement process: infrared light (IR) signal $x_I$, temperature $x_T$ measured by the NDIR sensor, and gas concentration $y$ measured by the co-located high-accuracy reference sensor. They are modeled as samples from real-valued random variables $X_I \in \mathbb{R}^+$, $X_T \in \mathbb{R}$, and $Y \in \mathbb{R}^+$ with a joint probability density function $p_{X_I,X_T,Y}(x_I, x_T, y)$.

There are two main sources of randomness that motivates the probabilistic modeling approach: measurement noise and unmeasured factors. For example, even when temperature and gas concentration are given, the IR signal remains random due to measurement noise and also unmeasured factors such as humidity, which affects the sensing process \cite{muller2020integration}. 

Over time, observations $(x_{I,i}, x_{T,i}, y_i)$, $i=1,\ldots,n$, are obtained and treated as independent samples from the joint distribution of $X_I$, $X_T$, and $Y$. This assumes that the temporal autocorrelations of the observations are neglected.

The sensor measurement model is given by
\begin{equation}
\label{estimator}
\widehat{Y} = f(X_T, X_I),
\end{equation}
where the estimator $\widehat{Y}$ is given by function $f(\cdot)$, which is based on a temperature-compensated Beer-Lambert law describing how chemical species absorb light \cite{swinehart1962beer}.\footnote{Note the difference between $\widehat{Y}$ and $Y$: Given $x_I$ and $x_T$, $\widehat{Y}$ is deterministic, while $Y$ is random, i.e., $Y | X_I, X_T \sim p_{Y | X_I, X_T}(y| x_I, x_T).$}

\subsection{Drift Analysis}

The root mean square error (RMSE) of the estimator $\widehat{Y}$ is

\begin{equation}
    \label{RMSE_E}
    \begin{split}
    &\text{RMSE}(\widehat{Y}) = \sqrt{\mathbb{E}[(\widehat{Y} - Y)^2]} \\
    &= \sqrt{\iiint ( f(x_I, x_T) - y)^2 p_{X_I,X_T,Y}(x_I, x_T, y) dx_I dx_T dy}.      
    \end{split}
\end{equation}

The RMSE of the sample is often used to empirically evaluate the sensor performance:
 \begin{equation}
 \label{RMSE_S}
     \text{RMSE} = \sqrt{\frac{1}{n}\sum_{i=1}^n \left(f(x_{T,i}, x_{I, i}) - y_{i}\right)^2}.
 \end{equation}

We use RMSE as the ``quantitative characteristic'' mentioned in Section \ref{intro} that may change over time. With the probabilistic approach, we can  decouple the sources of RMSE change.

In \eqref{RMSE_E}, the RMSE of the estimator is determined by the joint distribution if the measurement model $f(\cdot)$ is fixed. To isolate the different sources of drift, we need to factorize $p_{X_T,X_I,Y}(x_T, x_I, y)$. Based on the NDIR sensing principle, temperature $x_T$ and gas concentration $y$ cause the IR signal $x_I$, leading to the factorization:
\begin{equation}
p_{X_T,X_I,Y} = p_{X_T,Y}p_{X_I|X_T,Y}.
\end{equation}

Changes in both $p_{X_T,Y}$ and $p_{X_I|X_T,Y}$ change the joint density $p_{X_T,X_I,Y}$, affecting \eqref{RMSE_E} and \eqref{RMSE_S}. These changes have different sources: $p_{X_T,Y}$ depends on the sensor's environment, while $p_{X_I|X_T,Y}$ depends on the sensor's metrological properties. We define \textit{environmental variation} and \textit{instrumental drift} as temporal changes in $p_{X_T,Y}$ and $p_{X_I|X_T,Y}$, respectively. \footnote{Drift from unmeasured factors, like humidity, is included in instrumental drift as it adds randomness to \(p_{X_I|X_T,Y}\). If humidity is measured and included in the sensing model, it will be included in environmental variations.
}

Decoupling these sources is essential because environmental variation and instrumental drift have different characteristics and require different calibration techniques. Environmental variation is often caused by environmental dynamics or sensor relocation, and different conditions affect accuracy. Calibration for extreme conditions is unnecessary and may worsen sensor accuracy in normal operating conditions. In contrast, instrumental drift often evolves over time, requiring time-dependent models, such as autoregressive models.

In practice, environmental variation can be evaluated by studying samples of $X_T$ and $Y$. To isolate instrumental drift, we evaluate the expectation \eqref{RMSE_E} under the same density $p_{X_T,Y}$. Sampling methods can typically be applied to evaluate such expectations \cite[Ch~11]{bishop2006pattern}, which will be introduced next.

\subsection{Drift Evaluation}
\label{drift eva}
We use the importance sampling algorithm \cite{kloek1978bayesian} to resample from a given dataset according to a desired distribution. We then use \eqref{RMSE_S} to evaluate the RMSE and analyze the instrumental drift. Other sampling algorithms are covered in \cite{robert1999monte}.

\begin{algorithm}
\caption{Importance Sampling}
\begin{algorithmic}[1]
\State \textbf{Input:} Dataset $\{\mathbf{x}_i\}_{i=1}^N$, $P_{\text{desired}}(\mathbf{x})$, $P_{\text{original}}(\mathbf{x})$
\State \textbf{Output:} Resampled dataset $\{\mathbf{x}'_i\}_{i=1}^N$

\For{each $\mathbf{x}_i$ in the dataset}
    \State Calculate weight $w_i = \frac{P_{\text{desired}}(\mathbf{x}_i)}{P_{\text{original}}(\mathbf{x}_i)}$
\EndFor

\State Normalize weights $\tilde{w}_i = \frac{w_i}{\sum_{j=1}^N w_j}$

\State Resample $N$ data points from $\{\mathbf{x}_i\}_{i=1}^N$ according to $\tilde{w}_i$

\State \textbf{return} Resampled dataset $\{\mathbf{x}'_i\}_{i=1}^N$
\end{algorithmic}
\end{algorithm}

\section{Numerical Results}
Experiments were conducted using sensor data collected at the monitoring station in Dübendorf, Switzerland \cite{muller2020integration}, which include observational data from 18 low-cost Senseair LP8 NDIR CO$_2$ sensors and a co-located high-accuracy Picarro G1301 reference sensor, from 2017 to 2019. For LP8 sensors, automatic baseline correction (ABC) \cite{senseair2000} is turned off to study drift in its natural state.

Ten-minute averages of the measurements were evaluated. Due to missing values, data without common ten-minute measurements were removed. Six NDIR sensors had substantial missing data, so results are based on the remaining 12 sensors. Outliers in the NDIR sensor CO$_2$ readings, which were abnormally high due to water condensation \cite{muller2020integration}, were removed by retaining only CO$_2$ data within the 99.9\% percentile.

\begin{figure}[htbp]
\centerline{\includegraphics[width=0.4\textwidth]{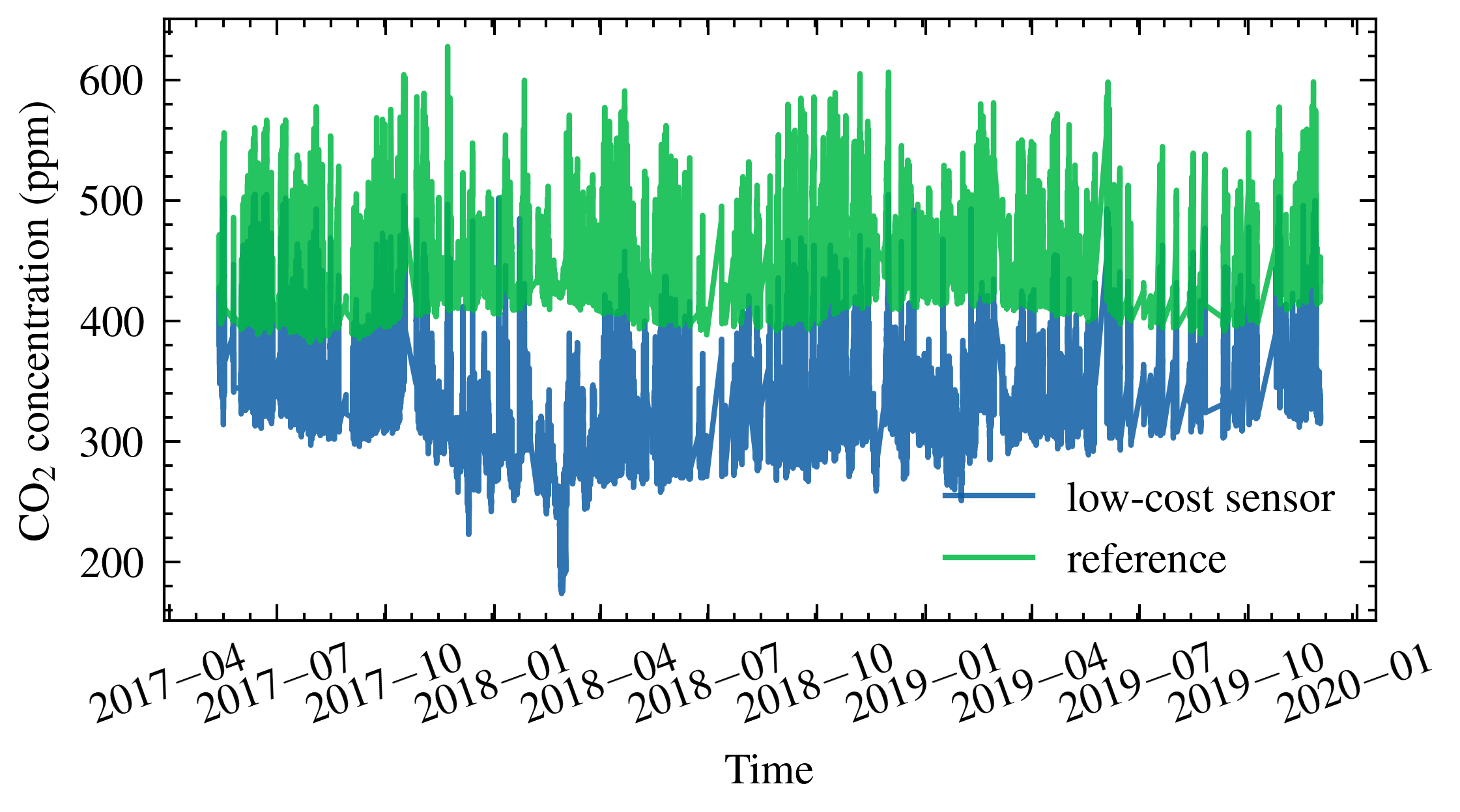}}
\caption{CO$_2$ measurements from low-cost sensor labeled \lb and the reference.}
\label{selected all}
\end{figure}

Fig.~\ref{selected all} shows CO$_2$ measurements from a low-cost sensor labeled \lb and the reference sensor as an example. The CO$_2$ measurements from this NDIR sensor are systematically lower than the reference and show performance drift over time.

\begin{figure}[htbp]
     \centering
     \begin{subfigure}[t]{0.4\textwidth}
         \centering
         \includegraphics[width=\textwidth]{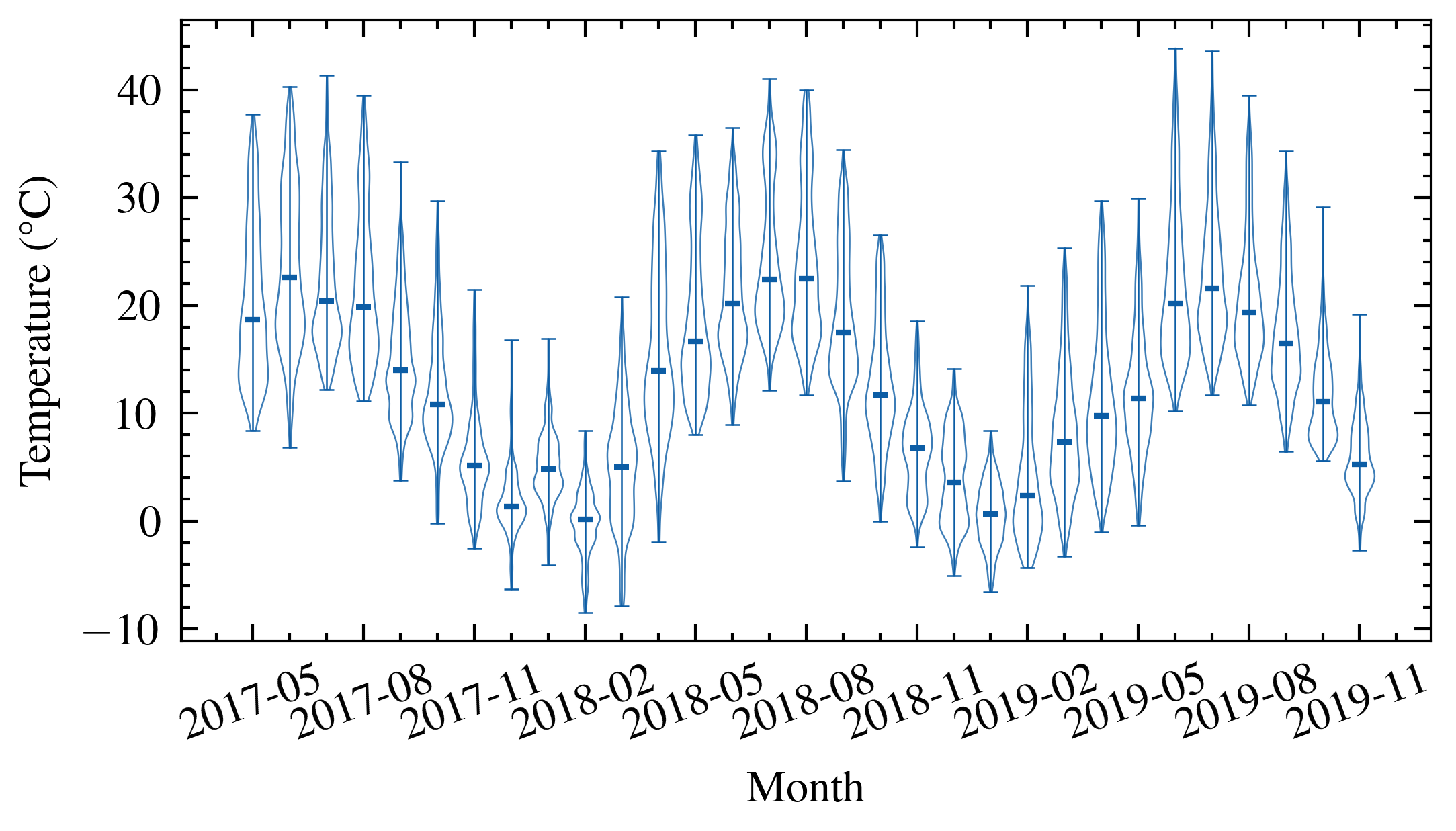}
         \caption{Violin plot of temperature measured by sensor \lb by month}
        \label{violin T}
     \end{subfigure}
     \hfill
     \begin{subfigure}[t]{0.4\textwidth}
         \centering
         \includegraphics[width=\textwidth]{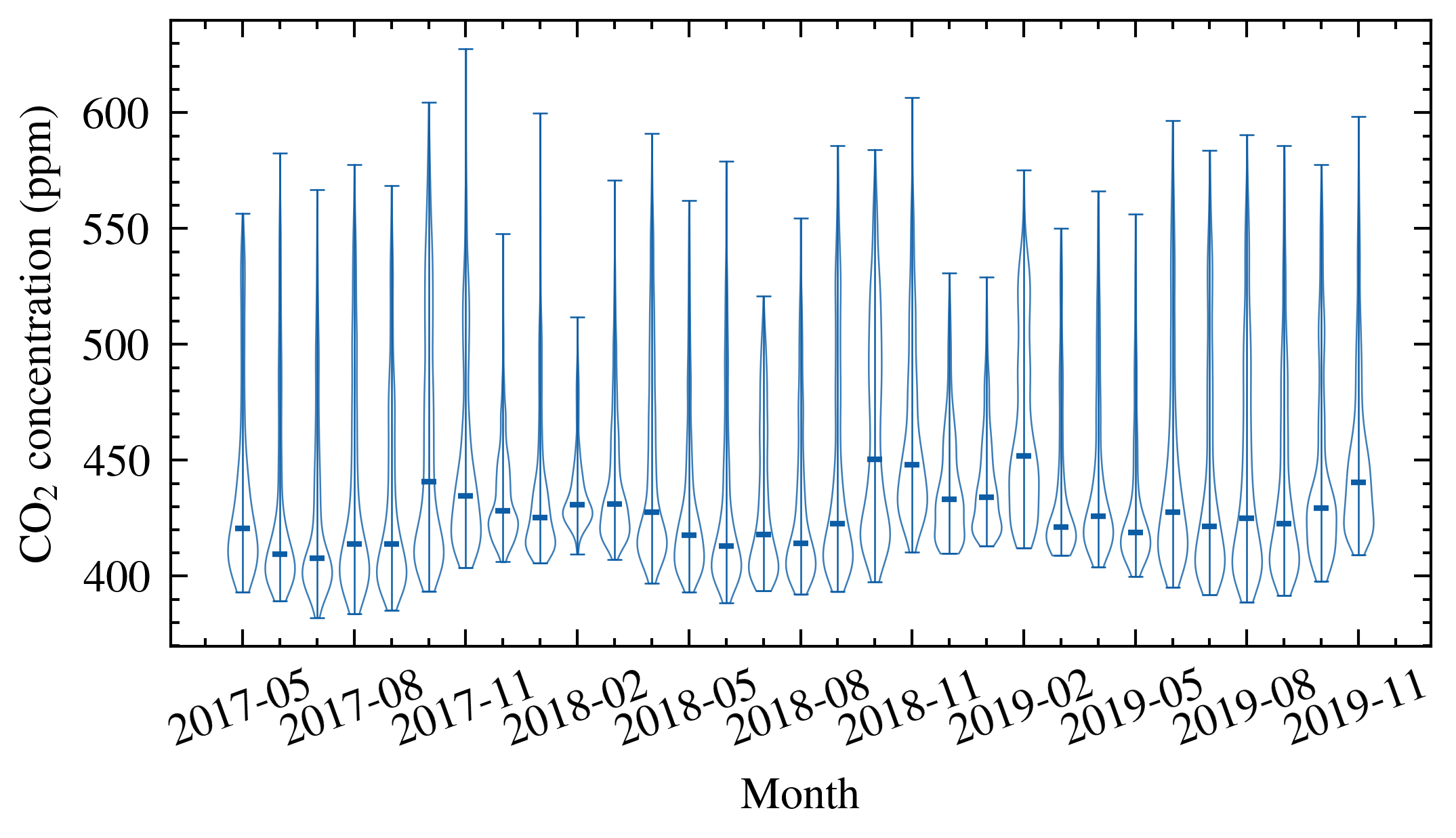}
        \caption{Violin plot of CO$_2$ measurement of the reference sensor by month.}
        \label{violin CO2}
     \end{subfigure}
     \caption{Evaluation of environmental variation.}
     \label{ev evaluation}
\end{figure}

We first evaluate environmental variation and then instrumental drifts. Due to space limits, we only show the marginal distribution changes of $X_T$ and $Y$ over time in Fig.~\ref{ev evaluation}.

\begin{figure}[htbp]
\centerline{\includegraphics[width=0.4\textwidth]{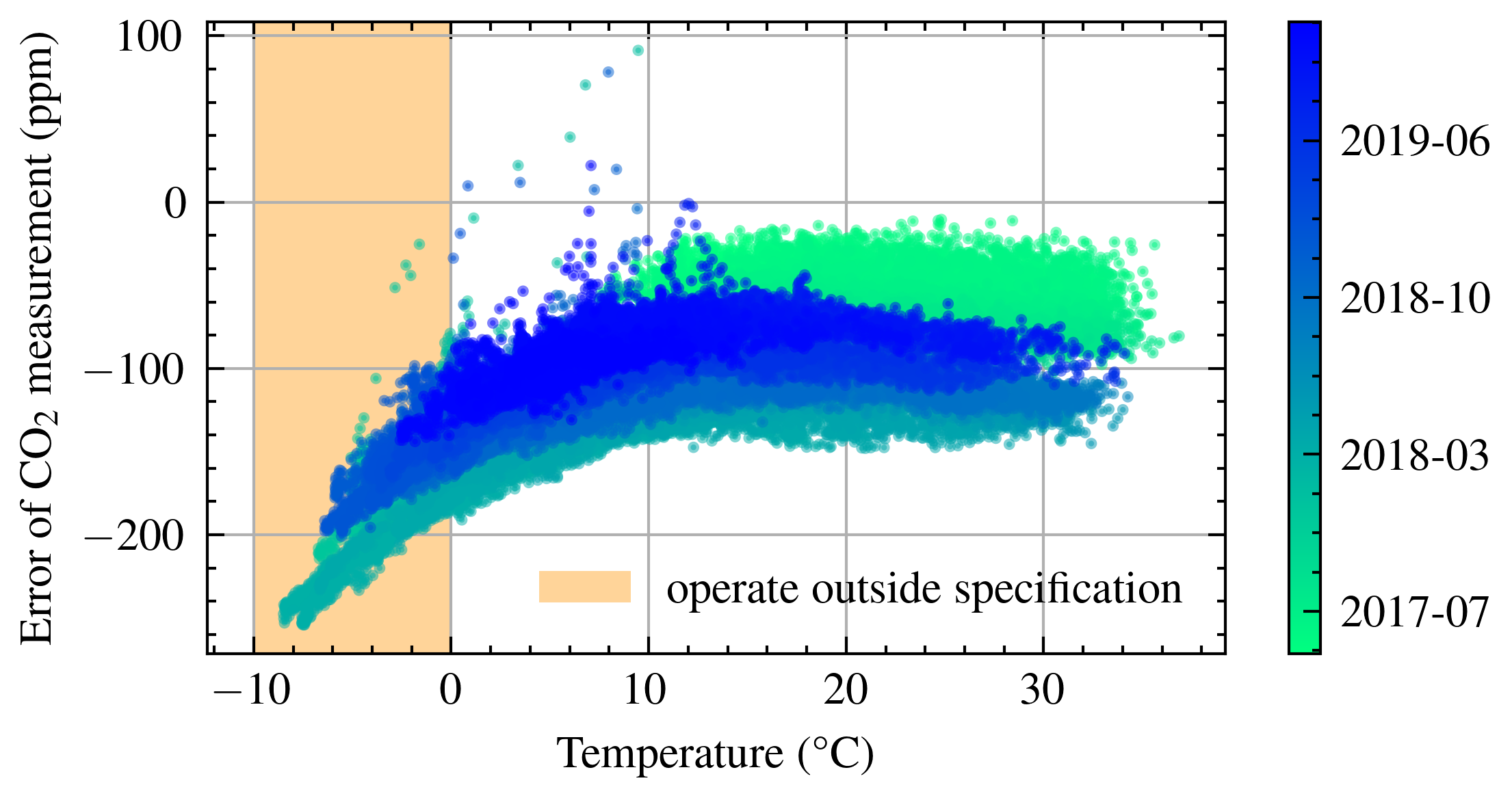}}
\caption{Scatter plot of CO$_2$ measurement error of sensor \lb and temperature, data points colored-coded by timestamp.}
\label{err vs T}
\end{figure}

Fig.~\ref{ev evaluation} shows violin plots \cite{hintze1998violin} by month of temperature measured by sensor \lb and CO$_2$ measured by the reference, which are often used to visually compare probability distributions. There are clear periodic patterns in the estimated temperature distributions, while the CO$_2$ distribution changes are relatively small. Short-term temperature variations (reflected by each vertical element of the violin plot) also impact the sensor's accuracy, as accuracy varies with temperature, shown in Fig.~\ref{err vs T}. Measurement error is systematically higher when temperature is out of the specified range $[0, 50]$ \si{\degreeCelsius}. We also see that the joint distribution of sensor error and temperature is stable in the short term but drifts over a longer period.


\vspace{-12pt} 

\begin{figure}[htbp]
\centerline{\includegraphics[width=0.4\textwidth]{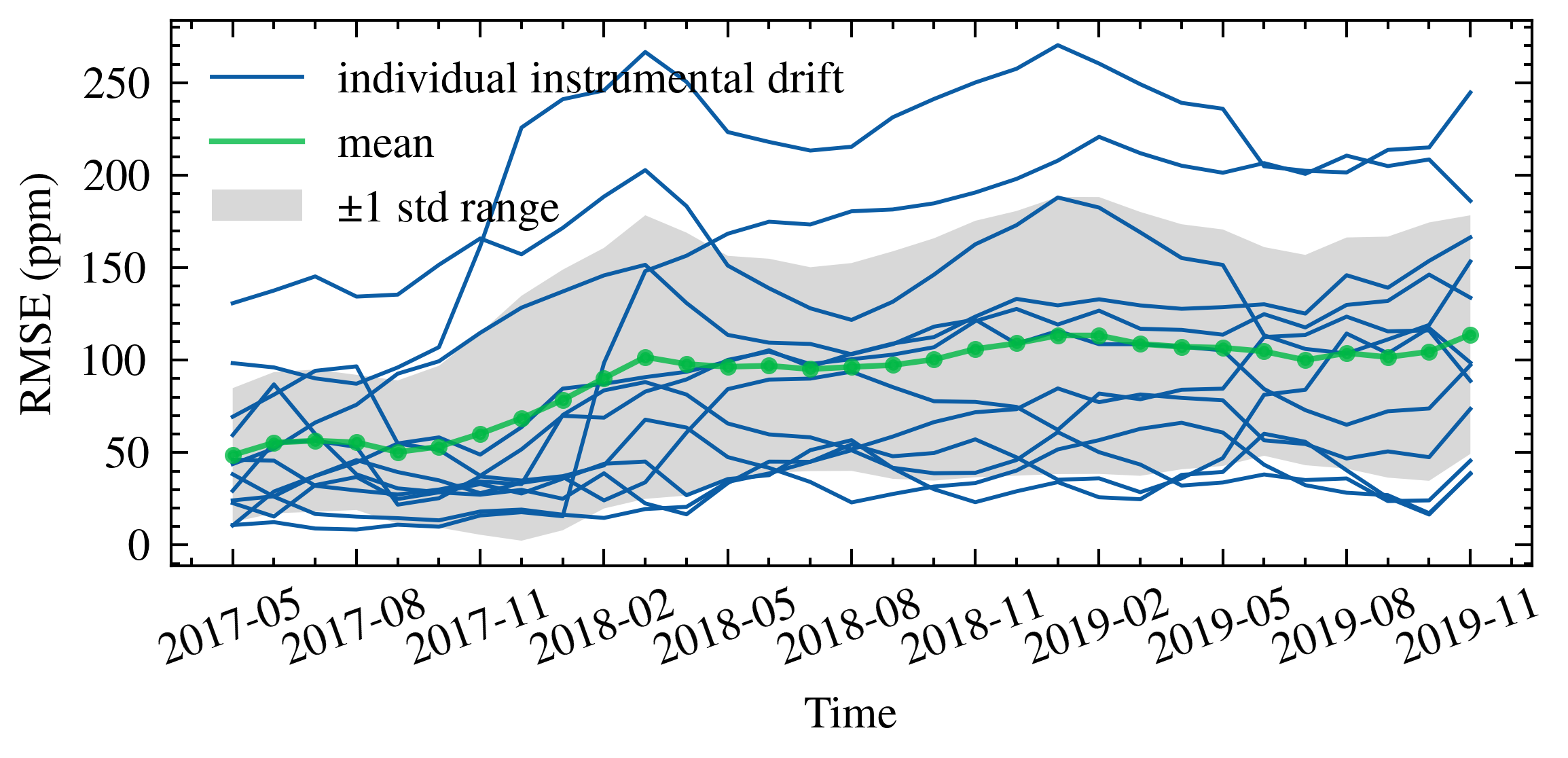}}
\caption{Instrumental drifts of 12 NDIR sensors by month, with mean and $\pm$ 1 std range.}
\label{RMSE all}
\end{figure}

\begin{figure}[htbp]
\centerline{\includegraphics[width=0.4\textwidth]{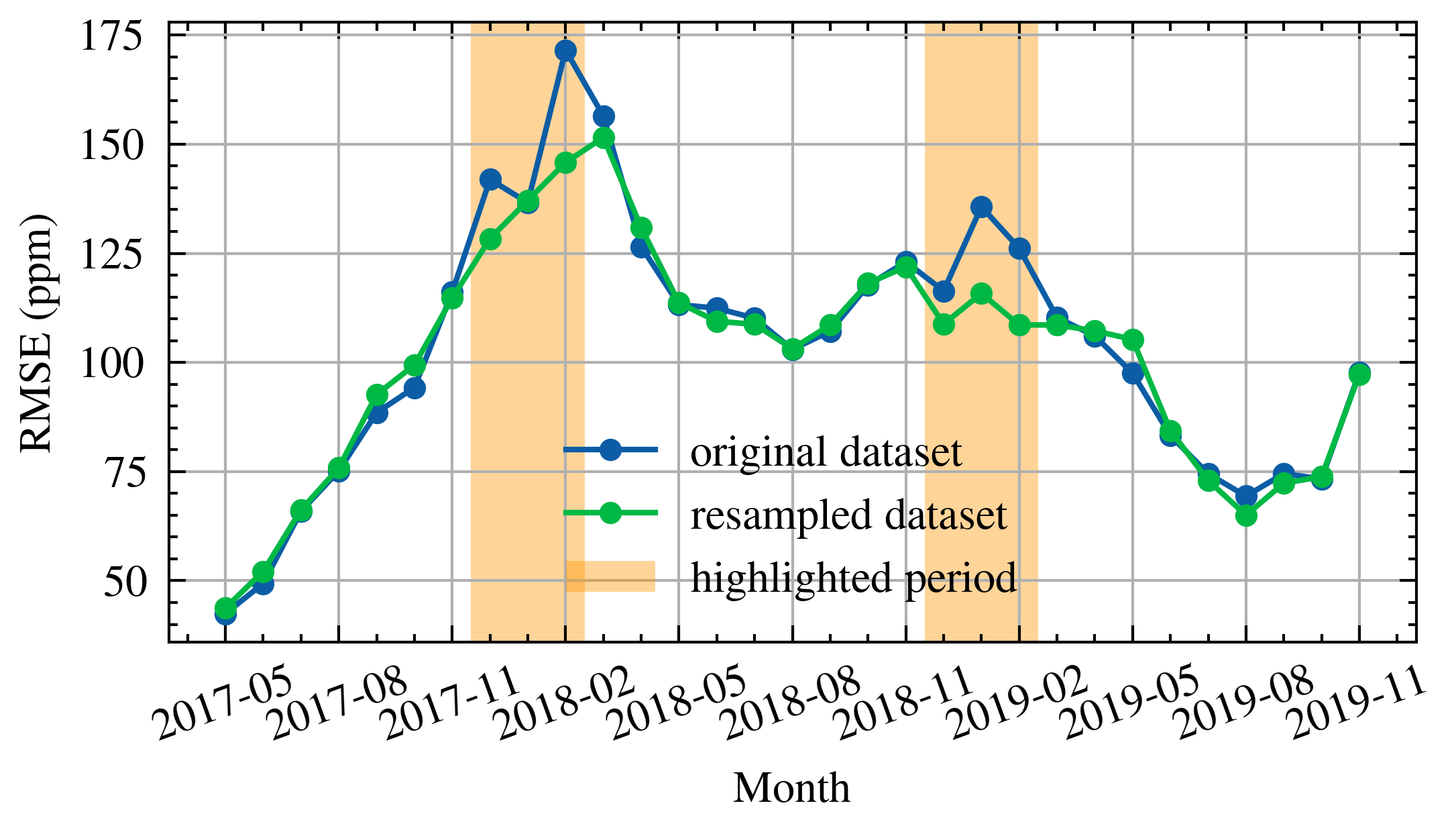}}
\caption{RMSE of sensor 1097, evaluated on the original dataset and the resampled dataset.}
\label{RMSE evaluation}
\end{figure}

To remove the effect of environmental variation and evaluate instrumental drift only, we applied the importance sampling algorithm in Section \ref{drift eva}. 500 measurements were resampled each month from the data according to the same density $p_{X_T,Y}$, a uniform distribution over $X_T \in [0, 20]\,\si{\degreeCelsius}$ and $y \in [400, 500]$ ppm. Fig.~\ref{RMSE all} shows the RMSE for 12 NDIR sensors by month with their mean and ±1 standard deviation range. The significant differences observed in individual drift patterns show the necessity of customized calibration for each sensor. RMSE of measurements from sensors \lb was evaluated on both the original and resampled datasets, shown in Fig.~\ref{RMSE evaluation}. RMSE values are similar for most months but differ during the two winter periods when temperature is lower, and $p_{X_T,Y}$ varies significantly from the sampling distribution.

\begin{figure}[htbp]
\centerline{\includegraphics[width=0.4\textwidth]{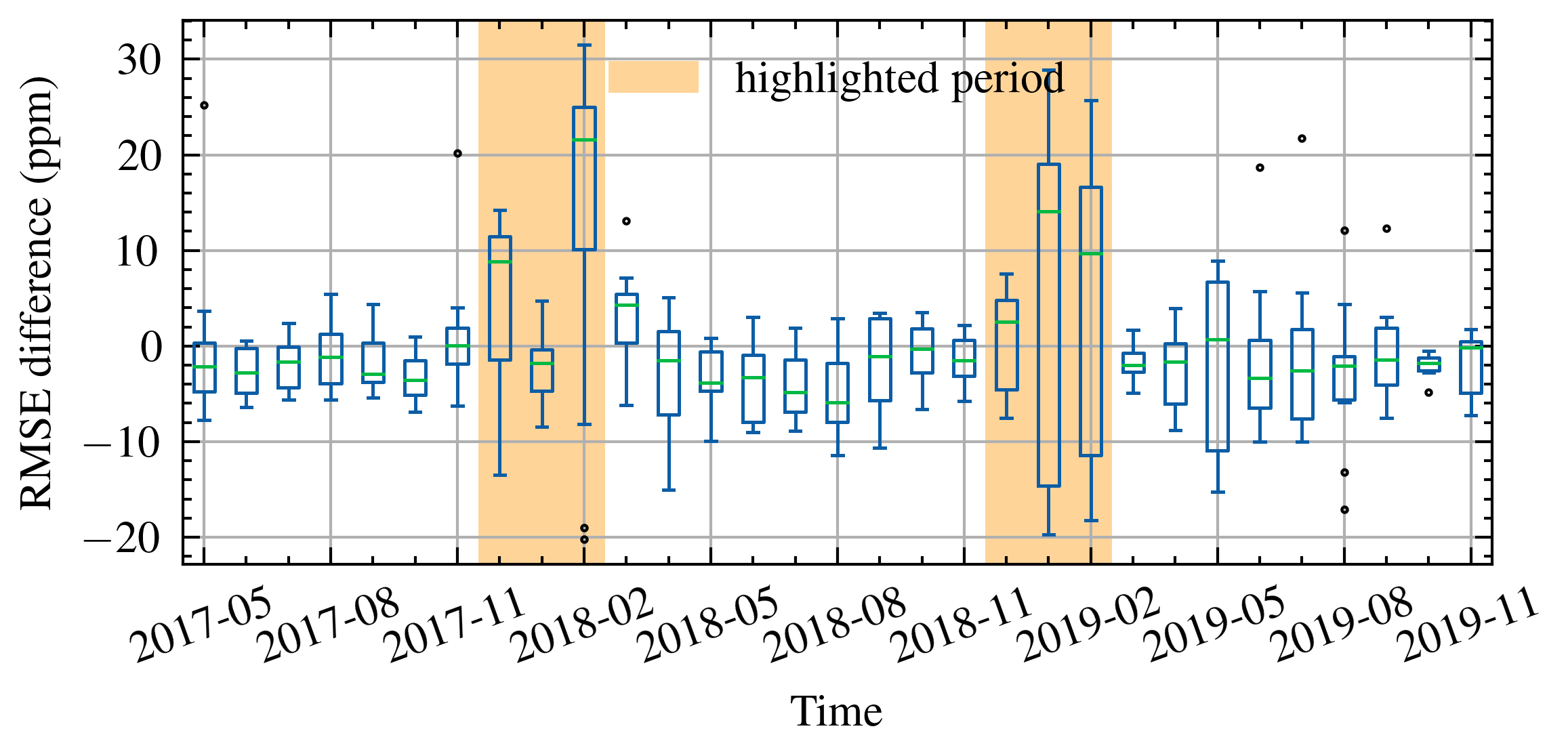}}
\caption{Box plot of the RMSE evaluation difference of 12 NDIR sensors by month.}
\label{box RMSE}
\end{figure}

Further, Fig.~\ref{box RMSE} shows the box plot of RMSE differences for all 12 sensors by month. During the two winter periods, RMSE differences have higher mean values and larger variation compared to other periods. The mean maximum absolute RMSE difference across sensors is 23.2 ppm, indicating that environmental variation significantly influences the evaluation of instrumental drift if not removed.

\vspace{-12pt} 
\vspace{-12pt} 

\section{Conclusion and Future Work}
We have presented a probabilistic approach to differentiate between environmental variation and instrumental drift, illustrated using NDIR CO$_2$ sensing. Through a detailed case study, we show that environmental conditions significantly influence sensor drift evaluation. By applying importance sampling, we effectively decouple these factors, leading to a more accurate assessment of instrumental drift. Future work will focus on further developing models for the drift process and exploring their application to the drift compensation.
\clearpage
\balance
\bibliography{refs}
\bibliographystyle{ieeetr}
\end{document}